\documentclass[pre,twocolumn,floatfix]{revtex4}
\usepackage[T1]{fontenc}
\usepackage[latin1]{inputenc}
\usepackage{amsmath}
\usepackage{amsfonts}
\usepackage{graphics}

\makeatletter

\usepackage{psfrag}

\makeatother

\begin{document}

\title{Electrons in 2D periodic magnetic and electric fields: 
Avoided band crossings, metal-insulator transitions, and
changes in the spreading direction}

\author{Bj\"orn Naundorf} 
\affiliation{Max-Planck-Institut f\"ur Str\"omungsforschung
and Institut f\"ur Nichtlineare Dynamik der Universit\"at G\"ottingen, \\
Bunsenstr.~10, 37073 G\"ottingen, Germany}

\author{Roland Ketzmerick}
\affiliation{Institut f\"ur Theoretische Physik,
Technische Universit\"at Dresden, 01062 Dresden, Germany}
\affiliation{Max-Planck-Institut f\"ur Str\"omungsforschung
and Institut f\"ur Nichtlineare Dynamik der Universit\"at G\"ottingen, \\
Bunsenstr.~10, 37073 G\"ottingen, Germany}

\date{\today}
\begin{abstract}
We study the energy spectrum and wave packet dynamics of electrons in a
two-dimensional periodic potential subject to a perpendicular
spatially modulated magnetic field. 
We observe avoided band crossings, metal-insulator transitions,
and changes in the spreading direction of wave packets
in two new situations: 
(i) for a purely magnetic modulation and 
(ii) in the limit of weak electric and magnetic modulations, 
where the Landau band coupling can be neglected.
In the second case the corresponding classical system is 
no longer chaotic but integrable, whereby 
these phenomena are not expected.
We argue that they are due to tunneling.
\end{abstract}

\pacs{73.43.Cd,73.40.Gk,05.45.Mt}
\maketitle

\section{Introduction}

Since the early days of quantum mechanics the problem of electrons in
a periodic potential subject to a magnetic field has been
studied \cite{peierlsetal}. 
While the energy spectrum of an electron in a
magnetic field is discrete, the spectrum of an electron in a
periodic potential has continuous energy bands.
The combined system
leads to fascinating commensurability effects, depending on the number of
magnetic flux quanta per unit cell of the potential. These effects are
visualized in the Hofstadter butterfly \cite{hofstadter} and are 
subject to intensive research.
While many studies focus on the case of uncoupled Landau
bands, which can be described by the Harper equation \cite{Harper},
the full system exhibits many more features
\cite{Schellnhuber1,Schellnhuber2,Schellnhuber3,GeiselPetschel,Kuehn1,Kuehn2}.
For example, unusual sequences of the Hall 
conductance appear~\cite{DennisPaper1}, which recently have been
observed experimentally with lateral superlattices on semiconductor
heterojunctions~\cite{Geisler}.
Another important feature of the Landau band coupling is
avoided band crossings, which occur due to a classically chaotic
limit and lead to metal-insulator transitions~\cite{KarstenPaper}. 
In systems with direction dependent modulation strengths 
they induce a counterintuitive 
change of the spreading direction of wave packets into the
direction of strong modulation~\cite{DennisPaper2}.
 
In the last few years attention has been paid to the
case of 2D periodically modulated {\em magnetic} fields
both in theory \cite{gerhardtspfannkuche,th1,th2,th3,th4,th5,th6,th7,th8,th9,th10,th11,th12,th13} 
and in experiment \cite{exp1,exp2,exp3,exp4}.
Consequently, the following question arises: Do avoided band crossings, metal-insulator transitions,
and the unusual change in transport direction
found for electric modulation~\cite{DennisPaper2} carry over to
magnetically modulated systems?

In this paper we derive a matrix equation for the energy spectrum of electrons
in 2D periodic electric and magnetic fields.
We observe avoided band crossings, metal-insulator transitions,
and changes in the spreading direction of wave packets
in two new situations:
Firstly, we find these phenomena in the case of a purely magnetic modulation.
Secondly, we study the limit of weak electric and magnetic modulations, 
where the Landau band coupling can be neglected.
For the energy spectrum we derive a 1D discrete Schr\"odinger equation 
with next-nearest neighbor coupling.
Its classical counterpart is an integrable 1D Hamiltonian, where
avoided band crossings are not expected.
We show, however, that avoided crossings exist and argue that they are due to
tunneling.
These avoided band crossings lead to metal-insulator transitions and 
changes in the spreading direction of wave packets in an unexpected regime. 

The paper is organized as follows:
We proceed in Sec.~II with the introduction of the system and show
in Sec.~III its spectrum and dynamics.
In Sec.~IV we study the limit of weak Landau band coupling.
The corresponding classical limit is shown to be integrable in
Sec.~V where we also discuss the origin of the avoided band
crossings. We finally summarize our results in Sec.~VI.

\section{System}
The dynamics of an electron with charge \( -e \) and mass \( m \) in a
potential \( V(\mathbf{r}) \) 
and a magnetic
field \( \mathbf{B}(\mathbf{r}) \) is described by the Hamiltonian
\begin{equation}\label{Eq:Hamiltonian}
H=\frac{1}{2m}\Bigl[\mathbf{p}+e\mathbf{A}(\mathbf{r})\Bigr]^{2}+V(\mathbf{r}),
\end{equation}
where the vector potential \( \mathbf{A}(\mathbf{r}) \) is given by 
\(\mathbf{B}(\mathbf{r})=\mathbf{\nabla }\times \mathbf{A}(\mathbf{r}) \).  
We are interested in a 2D potential 
and a perpendicular magnetic field, both periodic in
the \( x\) and \( y\) directions with period \( a \) and \( b
\), respectively. Both the potential and the magnetic field can be
decomposed into Fourier series,
\begin{equation} \label{Eq:PotFourierSum}
V(\mathbf{r})=\sum _{(r,s)\not =(0,0)}V_{r,s}e^{2\pi i(rx/a+sy/b)},
\end{equation}
\begin{equation} \label{Eq:BFourierSum}
\mathbf{B}(\mathbf{r})=\left( B_0+\sum _{(r,s)\not
=(0,0)}B_{r,s}e^{2\pi i(rx/a+sy/b)}\right) \mathbf{e}_{z}.
\end{equation}
We haven chosen the vector potential \( \mathbf{A}(\mathbf{r}) \) as,
\begin{widetext}
\begin{equation}
\mathbf{A}(\mathbf{r})=\left(0,B_0 x,0\right)+ \frac{1}{2\pi i}\sum
_{(r,s)\not =(0,0)}\frac{B_{r,s}}{
(r/a)^{2}+(s/b)^{2} }e^{2\pi i(rx/a+sy/b)} \left(
-s/b,r/a,0\right),
\end{equation}
\end{widetext}
such that it commutes with the momentum operator \(\mathbf{p}\), which 
reduces the following algebra. 
For vanishing magnetic and electric modulation, $B_{r,s}
\equiv V_{r,s} \equiv 0$, the eigenfunctions of $H$ are the well known
Landau level eigenfunctions,
 
\begin{equation}
\label{Eq:LandauEF}
\left<x,y| \nu ,\kappa \right\rangle =e^{i\kappa y/b} 
\varphi _{\nu }\left( \frac{x+\kappa l^{2}/b}{l}\right),
\end{equation}
with $\nu$ being the index of the harmonic oscillator eigenstate
$\varphi_\nu(z)=\exp(-z^2/2)H_\nu(z)$ \cite{CT}, 
the phase $\kappa \in \mathbb{R}$, and the magnetic length
$l^2=\hbar/(eB_0)$. In the presence of a modulation, the states
$\left|\nu,\kappa\right>$ are no longer eigenstates of $H$.
As they constitute a complete orthonormal set,
we use them as a convenient basis.

The magnetic flux $\Phi$ per unit cell of size $ab$ in
units of the magnetic flux quantum, $\Phi_0=h/e$, is
given by $\Phi/\Phi_0=abB_0/(h/e)$. If this ratio is rational,
$\Phi/\Phi_0=q/p$, the Hamiltonian commutes with the
magneto-translation operators
\begin{eqnarray}
M_{x+pa} & = & e^{ipay/l^2 }e^{pa\partial_x } \label{Eq:Magnetotrans1}\\
M_{y+b} & = & e^{b\partial_y }\label{Eq:Magnetotrans2},
\end{eqnarray}
and there are common eigenstates of $H$, $M_{x+pa}$, and $M_{y+b}$.
This will allow to treat the problem numerically. We note that while the 
set of rational magnetic fluxes \(\Phi/\Phi_0\) is of measure zero
within the real numbers, it is dense within the real numbers
and thus may approximate any irrational \(\Phi/\Phi_0\). 
The eigenequations of the magneto-translation operators read
\begin{eqnarray}
M_{x+pa} |\psi (k_{x},k_{y})\rangle & = & e^{2\pi i k_x} 
\left|\psi (k_{x},k_{y})\right>
\label{Eq:MagnetotransEigen1}\\
M_{y+b} \left|\psi (k_{x},k_{y})\right> & = & e^{2\pi i k_y}
\left|\psi (k_{x},k_{y})\right>
\label{Eq:MagnetotransEigen2},
\end{eqnarray} 
with the magnetic Bloch phases \(k_{x},k_{y}\in [0,1)\).
Their eigenvectors $\left|\psi (k_{x},k_{y})\right>$ can be constructed
from the chosen basis using
$M_{x+pa} \left|\nu,\kappa\right> = \left|\nu,\kappa+2\pi q\right>$
and
$M_{y+b} \left|\nu,\kappa\right>=e^{i\kappa}\left|\nu,\kappa\right>$,
yielding
\begin{equation}
\left|\psi (k_{x},k_{y})\right>=\sum _{\nu =0}^{\infty }\sum
^{\infty}_{n=-\infty}a_{\nu, n}(k_{x},k_{y}) 
\left| \nu ,2\pi (n+k_{y})\right\rangle,
\end{equation}
with 
\begin{equation}
a_{\nu, n+q}=e^{-2\pi i k_x} a_{\nu, n} \label{Eq:anunperiodisch}.
\end{equation}
Inserting $\left|\psi (k_{x},k_{y})\right>$ into the Schrödinger equation, 
$H\left|\psi(k_x,k_y)\right>=E\left|\psi(k_x,k_y)\right>$, and
projecting onto the basis states $\langle \nu,2\pi(n+k_y)|$ yields a set of
equations for the coefficients $a_{\nu,n}$,
\begin{widetext}
\begin{equation}\label{Eq:Matrixgleichung}
\left( \nu +\frac{1}{2}\right) a_{\nu, n}(k_{x},k_{y})
+\sum _{\nu '=0}^{\infty} \sum _{s=-\infty}^{\infty}
\mathbf{T}^{\nu ',\nu }_{n,s} a_{\nu ',n-s}(k_{x},k_{y})= 
 \frac{E}{\hbar \omega _{c}}a_{\nu, n}(k_{x},k_{y}),
\end{equation}
with the cyclotron frequency $\omega _{c}=eB_0/m$. 
This set can be reduced to values \( n=0,1,\ldots ,q-1 \) by employing
Eq.~(\ref{Eq:anunperiodisch}), which allows to write
$a_{\nu ',n-s}=
e^{-2\pi ik_{x}\left\lfloor (n-s)/q\right\rfloor } a_{\nu ', (n-s)\bmod q}$,
with \( \left\lfloor x\right\rfloor \)
denoting the largest integer that is smaller than or equal to \(x\). 
The coefficients $\mathbf{T}^{\nu ',\nu }_{n,s}$ are given by
\begin{eqnarray}
\mathbf{T}^{\nu ',\nu }_{n,s} =  
\mathop{\sum_{r}}_{(r,s)\not=(0,0)} &\Biggl[& \frac{V_{r,s}}{\hbar\omega_c} D^{r,s}_{\nu',\nu }
- \frac{1}{2\pi} \frac{q}{p} \mathop{\mathop{\sum_{r',s'}}_{(r',s')\not=(0,0)}}_{(r,s)\not=(r',s')} \frac{B_{r-r',s-s'}B_{r',s'}/B_0^2}
{(r-r')r'\alpha^2+(s-s')s'/\alpha^2}D_{\nu',\nu}^{r-r',s-s'} \\ 
& & -\frac{B_{r,s}/B_0}{2\pi(r^2+s^2/\alpha^4)} \sqrt{\frac{q }{p}} 
\left[ \sqrt{\nu '}D^{r,s}_{\nu '-1,\nu} (s\alpha+ir/\alpha)
      +\sqrt{\nu '+1}D^{r,s}_{\nu '+1,\nu } (-s\alpha+ir/\alpha)
\right] \Biggr] 
e^{-2\pi irp/q (n+k_{y}/2+s)}.\nonumber
\end{eqnarray}
\end{widetext}
 Here we have introduced \(\alpha=\sqrt{b/a}\) and
\begin{equation}
D^{\nu ',\nu }_{r,s}=e^{-\left| \rho \right| ^{2}/2}\sum _{k=0}^{\min
\{\nu ,\nu '\}}\frac{\sqrt{\nu '!\nu !}}{k!}\frac{\rho ^{\nu
'-k}(-\rho ^{*})^{\nu -k}}{(\nu '-k)!(\nu -k)!},
\end{equation}
where \( \rho =\sqrt{\frac{\pi p}{q}}(ir\alpha -s/\alpha ) \). 
The same result has been derived for $\alpha=1$ in a different way in
Ref.~\cite{gerhardtspfannkuche}.

In the following, we restrict ourselves to the lowest Fourier components in the
electric potential and the magnetic field as well as identical spatial periods \( a=b
\) such that,
\begin{equation}
\label{PotLowFourier}
V(x,y)=\hbar \omega _{c}\left( v_{x}\cos \frac{2\pi x}{a}+v_{y}\cos \frac{2\pi y}{a}\right) ,
\end{equation}
\begin{equation}
\label{MagLowFourier}
\mathbf{B}(x,y)=B_0\left( 1+b_{x}\cos \frac{2\pi x}{a}+b_{y}\cos \frac{2\pi x}{a}\right) \mathbf{e}_{z},
\end{equation}
i.e.~in the notation of Eqs.~(\ref{Eq:PotFourierSum}) and
(\ref{Eq:BFourierSum}), \( B_{-1,0}=B_{1,0}=B_0 b_{x}/2 \), \(
B_{0,-1}=B_{0,1}=B_0 b_{y}/2 \), \(
V_{-1,0}=V_{1,0}=\hbar \omega_{c} v_{x}/2  \) and \(
V_{0,-1}=V_{0,1}=\hbar \omega_{c} v_{y}/2  \).

\section{Spectrum and Dynamics}

%%%%%%%%%%%%%%%%%%%%%%%%%%%%%%%%%%% Fig 1 %%%%%%%%%%%%%%%%%%%%%%%%%%%%%%%%%%%%
\begin{figure*}
\psfrag{X}[c]{\large $b_x+b_y$}\psfrag{Energy}[c]{\large $E/\hbar\omega_c$}
{\centering \resizebox*{
\textwidth}{!}{\includegraphics{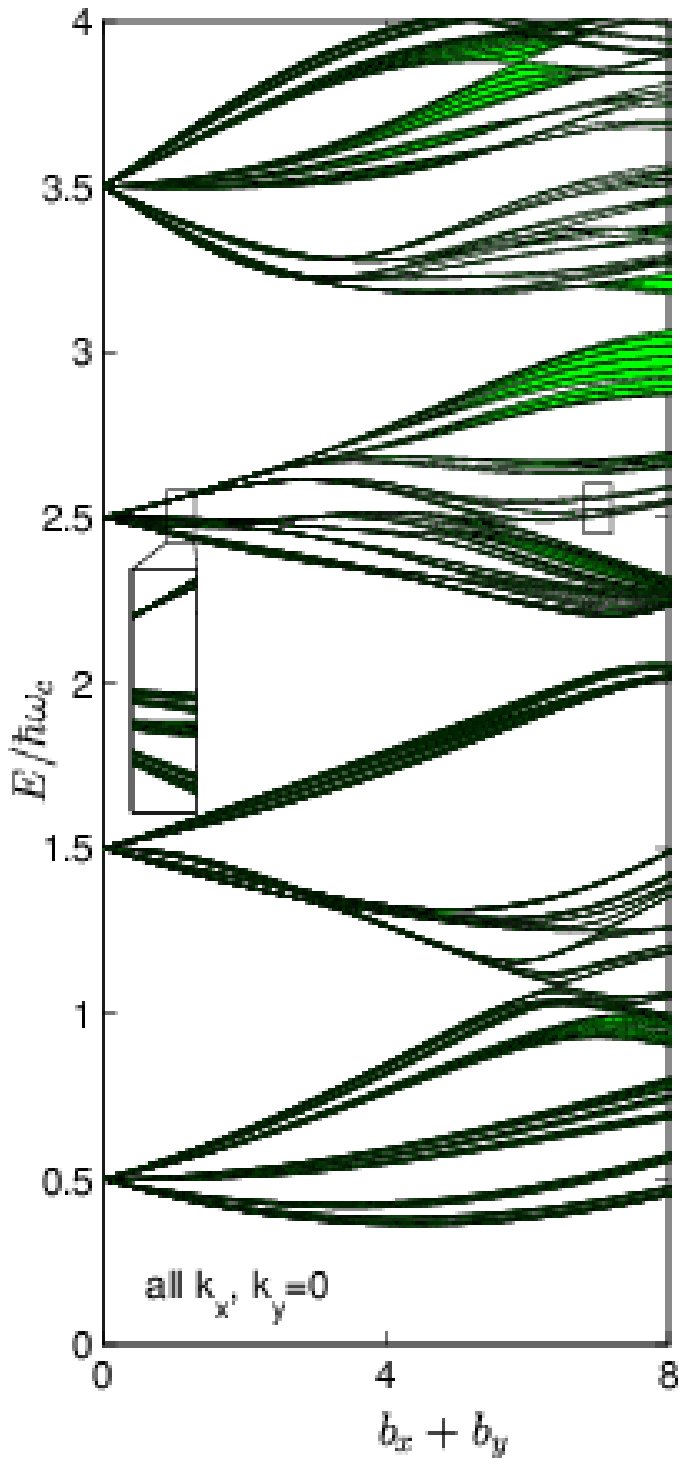}
\includegraphics{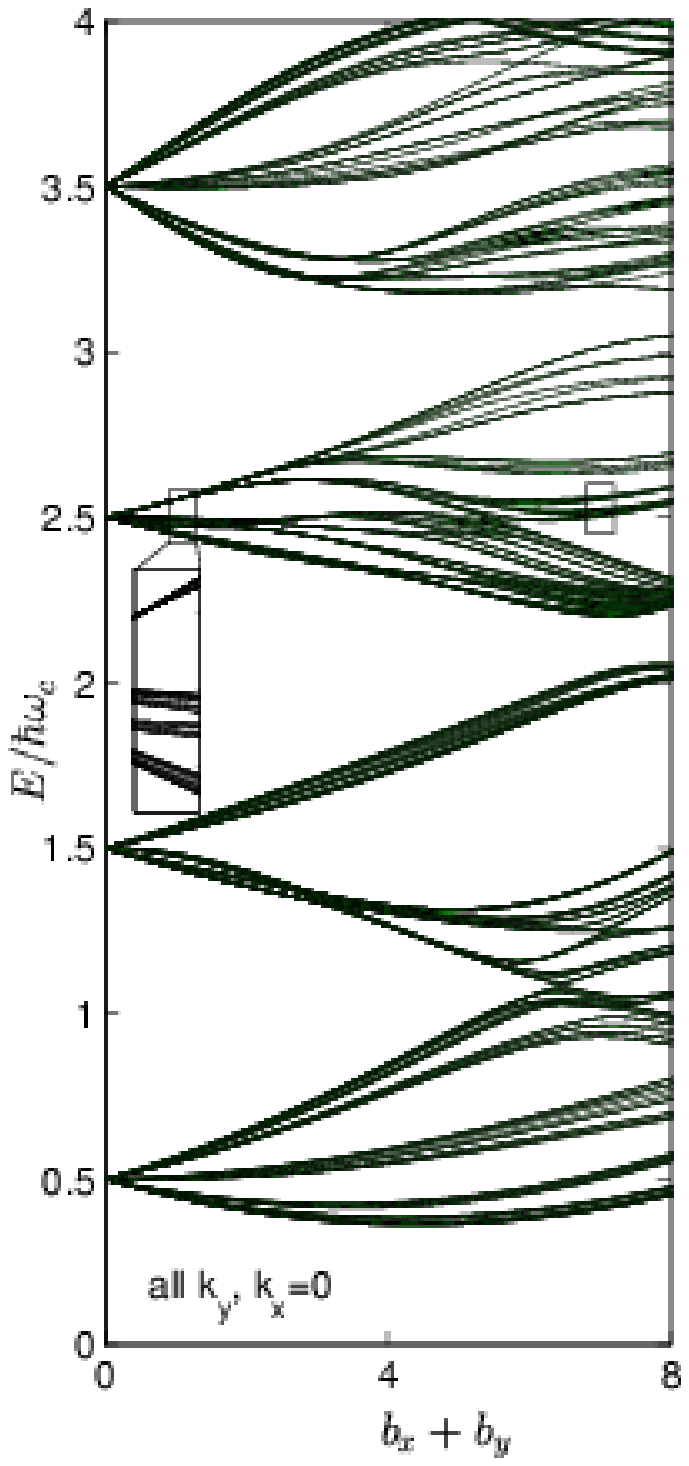}} \par}

\caption{\label{Abb:Energiespektrumkxky}
Spectrum of Eq.~(\ref{Eq:Hamiltonian})
vs. the strength of the magnetic modulation, $b_x+b_y$,
for a fixed ratio of modulation strengths, $b_x/b_y=0.8$,
and without electrostatic modulation, $v_x=v_y=0$.
The average magnetic flux per unit cell is given by
$\Phi/\Phi_0=55/89$, which is an approximant of the golden mean
\((\sqrt5-1)/2\).
In the left part, \protect\( k_{x}\protect \) is
varied, keeping \protect\( k_{y}=0\protect \) fixed, 
while in the right
part \protect\( k_{y}\protect \) is varied, keeping \protect\(
k_{x}=0\protect \) fixed. 
The lowest four Landau bands are shown, while eight Landau bands were taken
into account for the numerical calculation.
The boxes depict the energy interval chosen for the wave packet dynamics in
Fig.~\ref{Abb:Spreading}.
}
\end{figure*}
%%%%%%%%%%%%%%%%%%%%%%%%%%%%%%%%%%%%%%%%%%%%%%%%%%%%%%%%%%%%%%%%%%%%%%%%%%%%%%%

Figure \ref{Abb:Energiespektrumkxky} shows the energy spectrum
for a purely magnetic modulation, \(v_x=v_y=0\),
and for different magnetic modulation strengths in the
$x$ and $y$ directions, \(b_x/b_y=0.8\).
One can see how the isolated Landau levels at \(b_x=b_y=0\) 
are split up into a complex subband structure for \(b_x+b_y>0\). 
Upon further increase of the magnetic modulation strength, 
these bands broaden but typically repel and do not cross each other. 
This phenomenon has also been termed \emph{avoided band crossings}
\cite{KarstenPaper}. 
In the left panel a subspectrum is shown where
only $k_x$ is varied and $k_y=0$ is kept
fixed, whereas in the right part $k_y$ is varied, keeping $k_x=0$ fixed.
A clear duality of these subspectra can be seen with wide bands in one panel
corresponding to narrow bands in the other panel.
In the limiting case of irrational $\Phi/\Phi_0$ this corresponds to
absolutely continuous and pure point spectral regions, respectively.
Therefore, even in the rational case, where we present the numerics, 
we will denote the narrow bands as levels.

\begin{figure}
{\par\centering \psfrag{x }[c]{\large $x$}
  \psfrag{y }[c]{\large $y$}\psfrag{a}[c]{\large $(a)$}
  \psfrag{b}[c]{\large $(b)$}
  \psfrag{t}[c]{\large time}
  \resizebox*{0.5 \textwidth}{!}{\includegraphics{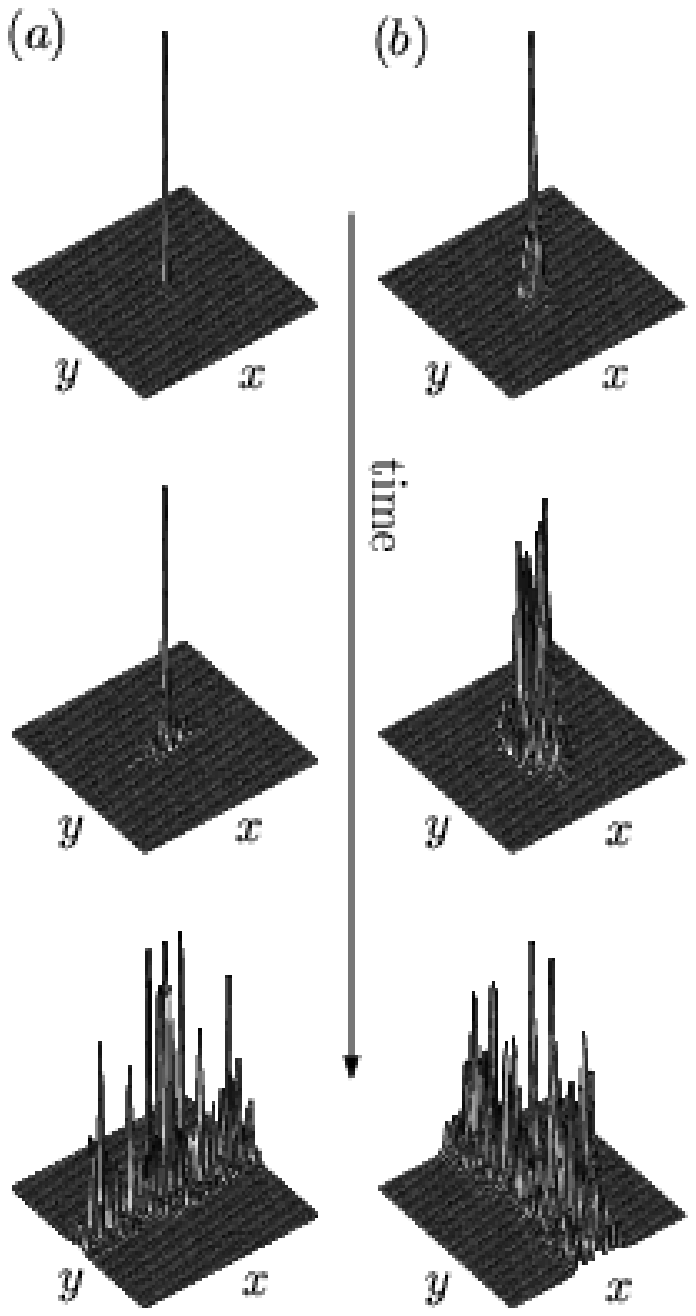}} \par}
 \caption{\label{Abb:Spreading} Spreading of initially localized
   wave packets for magnetic modulation strengths
   (a) $b_x+b_y=1.1$ and (b) $b_x+b_y=7.0$
   at a fixed ratio $b_x/b_y=0.8$ and $v_x=v_y=0$.
   The wave packets
   are constructed from the eigenstates in the energy intervals
   depicted in Fig.~\ref{Abb:Energiespektrumkxky}.
 }
\end{figure}

Such spectral properties go along with directional properties of eigenstates, 
which was demonstrated for a purely electrostatic modulation in 
Ref.~\cite{DennisPaper2}.
For example, a band under variation of $k_x$ and a level under 
variation of $k_y$ correspond to eigenfunctions extended in the $x$
direction and localized in the $y$ direction.
This has immediate consequences for the dynamics of wave packets.
An initially localized wave packet, constructed from states in an 
energy interval with bands under variation of $k_x$, will 
spread ballistically in the $x$ direction and
will localize in the $y$ direction.
This is demonstrated in Fig.~\ref{Abb:Spreading}a.
The initial wave packet is constructed from a symmetric eigenstate of the
$\nu=2$ Landau level that is projected
on the subspace spanned by the eigenstates of the chosen energy range.

%%%%%%%%%%%%%%%%%%%%%%%%%%%%%%%%%%% Fig 2 %%%%%%%%%%%%%%%%%%%%%%%%%%%%%%%%%%%%
%%%%%%%%%%%%%%%%%%%%%%%%%%%%%%%%%%%%%%%%%%%%%%%%%%%%%%%%%%%%%%%%%%%%%%%%%%%%%%%

In Fig.~\ref{Abb:Spreading}b we present the spreading behavior for
stronger magnetic 
modulation at the same ratio $b_x/b_y=0.8$ and for a similar energy range.
We find that the spreading direction is switched to the opposite direction.
This nonintuitive dynamical effect was previously observed in the case
of a purely electrostatic modulation~\cite{DennisPaper2}.
It can be understood from the subspectra shown in 
Fig.~\ref{Abb:Energiespektrumkxky}.
Upon the increase of the magnetic modulation strength $b_x+b_y$
one observes metal-insulator transitions in one subspectrum that go along with
insulator-metal transitions in the dual subspectrum.
In Ref.~\cite{KarstenPaper} it was argued that such metal-insulator
transitions can be induced by avoided band crossings, which
are ubiquitous in the spectrum of 
Fig.~\ref{Abb:Energiespektrumkxky}.

Avoided band crossings and avoided level crossings are a generic 
phenomenon in quantum systems with a
classically chaotic limit.
In fact, the present system with magnetic modulation was shown to be 
chaotic~\cite{Schmelcher}. 
Therefore the observed change in the transport direction by increasing the 
magnetic modulation strength can be understood along the line of 
reasoning previously applied to the case of a purely electrostatic modulation:
A classically chaotic limit leads to avoided band crossings in extended 
systems. They may induce metal-insulator transitions~\cite{KarstenPaper}
which lead to a change in transport direction~\cite{DennisPaper2}.
In addition to these findings for a purely magnetic modulation
we have verified that such changes in the transport direction just as well
occur in the general case of combined electric and magnetic modulation.
  
In the second part of the paper we want to demonstrate that there is 
a second origin of avoided band crossings that leads to the same
spectral and dynamical phenomena.
To this end, we will study a limit where the classical dynamics
is not chaotic and where we will argue that avoided 
crossings are due to tunneling.

\section{The limit of negligible Landau band coupling}

For small modulations strengths $b_x, b_y, v_x, v_y \ll 1$
we can neglect the coupling between Landau bands. 
Thus we ignore matrix elements with different \( \nu \) and \( \nu '\),
leading to a discrete 1D Schr\"odinger equation
with next-nearest neighbor coupling for every Landau band,
\begin{widetext}
\begin{equation}
\label{Harper2}
t_{1}(a_{n-1}+a_{n+1})+t_{2}(a_{n-2}+a_{n+2}) +
\left[ 2U_{1}\cos \left( 2\pi \frac{\Phi _{0}}{\Phi }(n+k_{y})\right)
+2U_{2}\cos \left( 4\pi \frac{\Phi _{0}}{\Phi }(n+k_{y})\right) \right] a_{n}
=\tilde{E}a_{n},
\end{equation}
\end{widetext}
with $a_{n+q}=e^{-2\pi i k_x} a_n$.
The parameters $U_1$, $U_2$, $t_1$, and $t_2$ are given by the magnetic
and electrostatic modulation strengths,
\begin{eqnarray}
U_{1} & = & b_{x}s_{1}+v_{x}s_{2}\\
t_{1} & = & b_{y}s_{1}+v_{y}s_{2}\\
U_{2} & = & -b_{x}^{2}s_{3}\\
t_{2} & = & -b_{y}^{2}s_{3},
\end{eqnarray}
and \( \tilde{E}=E/\hbar \omega _{c}-\left[ \left( \nu +\frac{1}{2}\right) +(\Phi/\Phi_0)\left( b_{x}^{2}+b_{y}^{2}\right)/(8\pi) \right]  \).
Here we have introduced the following coefficients, 
\begin{eqnarray}
s_{1} & = & \sqrt{\nu }\sum _{k=0}^{\nu -1}\frac{(\pi \Phi _{0}/\Phi )^{\nu -1-k}}{(\nu -k)!(\nu -1-k)!}\\ \nonumber
&&+\sqrt{\nu +1}\sum _{k=0}^{\nu }\frac{(\pi \Phi _{0}/\Phi )^{\nu -k}}{(\nu -k)!(\nu +1-k)!}\\
s_{2} & = & \sum _{k=0}^{\nu }\frac{\nu !}{k!}\frac{(\pi \Phi _{0}/\Phi )^{\nu -k}}{(\nu -k)!^{2}}\\
s_{3} & = & e^{-\frac{3}{2}\pi \frac{\Phi _{0}}{\Phi }}\sum _{k=0}^{\nu }\frac{\nu !}{k!}\frac{(-4\pi \Phi _{0}/\Phi )^{\nu -k}}{(\nu -k)!^{2}},
\end{eqnarray}
which solely depend on the Landau level index \( \nu  \) and the magnetic
flux \( \Phi /\Phi _{0} \).

\begin{figure}
{\centering \psfrag{x}[c]{\large $U_1$}\psfrag{y}[c]{\large $\tilde
E$}\psfrag{A}[x]{\large $\left|a_n\right|^2$}
\psfrag{n}[c]{\large $n$}\resizebox*{0.48
\textwidth}{!}{\includegraphics{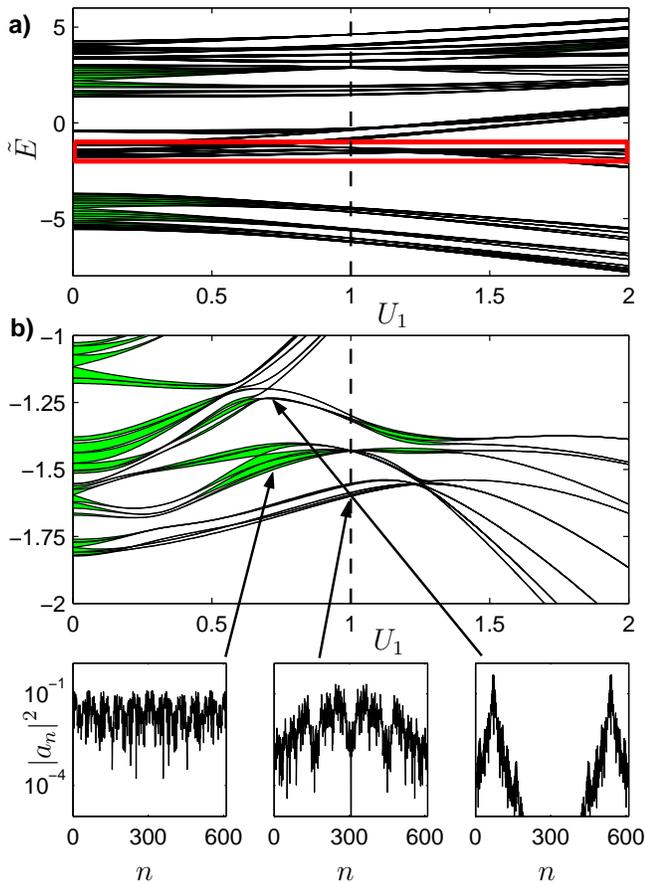}} \par}
\caption{\label{Abb:Harper2EMU1}
 Spectrum of Eq.~(\ref{Harper2}) vs.~\protect\( U_{1}\protect \)
 under variation of \protect\( k_{x}\protect \) with $q/p=377/610$ 
 for \protect\( t_{1}=1\protect \), \protect\(
 t_{2}=1.5\protect \), and \protect\( U_{2}=1.5\protect \) (a) and
magnification of the boxed region (b). At the selfdual point $U_1=1$
all eigenstates are fractal. Many avoided band crossings can be
seen and there are parameters values for which extended (bottom left) and
localized eigenstates (bottom right) coexist.
}
 \end{figure}

In the special case $t_2=U_2=0$, i.e. without magnetic modulation,
Eq.~(\ref{Harper2}) corresponds to the Harper equation \cite{Harper}.
In the general case, an important property of the Harper
equation carries over: the Aubry duality~\cite{AubryAndre}.
In order to see this, we multiply Eq.~(\ref{Harper2}) by
\( e^{2\pi i(n+k_{y})m\Phi _{0}/\Phi } \) and sum over \(n\).
With \( b_{m}=\sum _{n=-\infty }^{\infty }e^{2\pi i(n+k_{x})m\Phi _{0}/\Phi }a_{n} \)
this leads to a dual equation
\begin{widetext}
\begin{equation}
U_{1}(b_{m-1}+b_{m+1})+ U_{2}(b_{m-2}+b_{m+2}) +
\left[ 2t_{1}\cos \left( 2\pi \frac{\Phi _{0}}{\Phi}(m+k_{x})\right)
+2t_{2}\cos \left( 4\pi \frac{\Phi _{0}}{\Phi}(m+k_{x})\right) \right] b_{m}
=\tilde{E}b_{m},
\end{equation}
\end{widetext}
where $U_1$, $U_2$ are exchanged with $t_1$, $t_2$.
The selfdual point is given by \( U_{1}=t_{1} \), \( U_{2}=t_{2}\),
i.e.\ when the modulation strengths are equal in the
$x$ and $y$ directions, $v_x=v_y$ and $b_x=b_y$.
Like in the Harper equation one expects here a fractal spectrum
and fractal eigenfunctions, as visualized in Fig.~\ref{Abb:Harper2EMU1}.
The duality implies that from an eigenstate at one parameter set one can
directly construct the eigenstate at the dual parameter set.
In particular, a localized eigenstate leads to a dual state that is
extended~\cite{AubryAndre}.

%%%%%%%%%%%%%%%%%%%%%%%%%% Fig 3 %%%%%%%%%%%%%%%%%%%%%%%%%%%%%%%%%%%%%
%%%%%%%%%%%%%%%%%%%%%%%%%%%%%%%%%%%%%%%%%%%%%%%%%%%%%%%%%%%%%%%%%%%%%%%

An important property of the Harper equation, that for
$U_1/t_1<1$ all eigenstates are extended and for
$U_2/t_2>1$ all eigenstates are localized (at least for most irrational
$\Phi/\Phi_0$) does not carry over.
In fact, we see in Fig.~\ref{Abb:Harper2EMU1} that for a given parameter set
localized and extended eigenstates coexist.
This is accompanied by avoided band crossings and thus can be understood
on the basis of the analysis in Ref.~\cite{KarstenPaper}.

\begin{figure}
{\centering \psfrag{x}[c]{\large $t_2$}\psfrag{y}[c]{\large $\tilde
E$}\psfrag{n}[c]{\large $n$}\psfrag{A}[c]{\large$\left|a_n\right|^2$}\resizebox*{0.5
\textwidth}{!}{\includegraphics{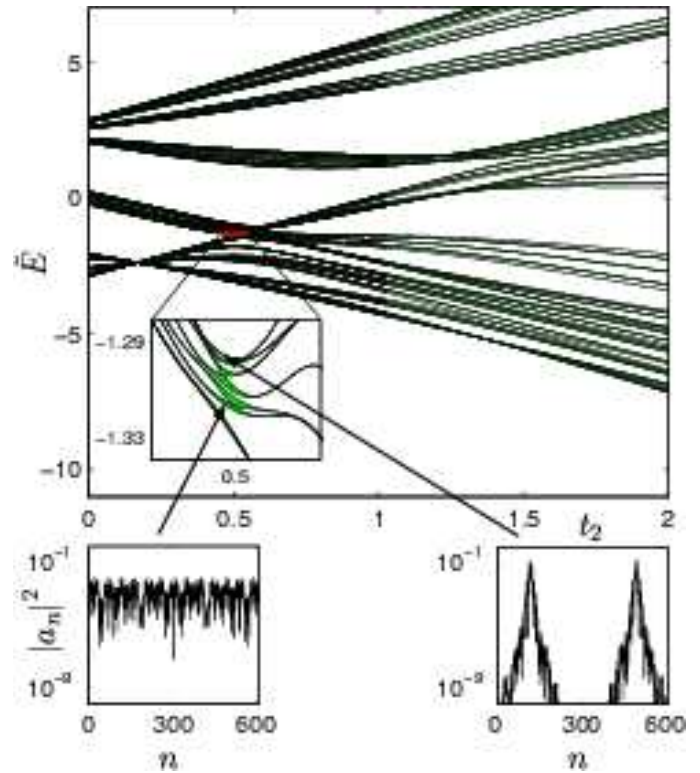}} \par}
\caption{\label{Abb:Harper2U1}
 Spectrum of Eq.~(\ref{Harper2}) vs.~\protect\( t_{2}\protect \)
 for a purely magnetic modulation,
 \protect\( \left( U_{1}/t_{1}\right) ^{2}=U_{2}/t_{2}\protect \),
 under variation of \protect\( k_{x}\protect \) with $q/p=377/610$.
 Further parameters are \protect\( t_{1}=1\protect \) and
 \protect\( U_{1}=1.2\protect \).
 The inset shows avoided band crossings and the coexistence of
 extended and localized eigenstates.
}
\end{figure}

Two questions arise at this point:
Firstly, one might wonder if in the special case of a purely magnetic modulation,
like for a purely electrostatic modulation (Harper equation), all eigenstates for a
given parameter set are either localized or extended.
The answer is given by Fig.~\ref{Abb:Harper2U1}, which
shows an energy spectrum for the case of
purely magnetic modulation, i.e.\ (\( U_{2}/t_{2})=(U_{1}/t_{1})^{2} \).
We still observe avoided band crossings and the coexistence of localized
and extended states.
We conclude that the existence of the terms $t_2$ and $U_2$
in Eq.~(\ref{Harper2}) leads to this deviation from the qualitative
properties of the Harper equation.
As an aside, we note that we have to zoom into finer details of the
spectrum to see this effect in the case of a purely magnetic modulation.

%%%%%%%%%%%%%%%%%%%%%%%%% Fig 4 %%%%%%%%%%%%%%%%%%%%%%%%%%%%%%%%%%%%%%%
%%%%%%%%%%%%%%%%%%%%%%%%%%%%%%%%%%%%%%%%%%%%%%%%%%%%%%%%%%%%%%%%%%%%%%%%

The second and more important question stems from the fact that
discrete 1D Schr\"odinger equations like Eq.~(\ref{Harper2})
have a classical limit that is integrable (see below),
while it was previously argued~\cite{KarstenPaper, DennisPaper2}
that avoided band crossings are induced by a classically chaotic limit:
Why do we observe avoided band crossings in this classically non-chaotic case?

\section{Classical limit, Tunneling, and Avoided Band Crossings}

A 1D discrete Schr\"odinger equation
can be written in a Hamiltonian form:
Equation~(\ref{Harper2}) with
\( x=2\pi \frac{\Phi _{0}}{\Phi }(n+k_{y}) \)
and \( \psi (x)=a_{n} \) can be written as, 
\begin{widetext}
\begin{equation}
t_{1}\bigl[ \psi( x-2\pi \frac{\Phi _{0}}{\Phi })
+ \psi( x+2\pi \frac{\Phi _{0}}{\Phi }) \bigr]
+ t_{2}\bigl[ \psi( x-4\pi \frac{\Phi _{0}}{\Phi })
+ \psi( x+4\pi \frac{\Phi _{0}}{\Phi }) \bigr]
+ 2\bigl[U_{1}\cos \left( x\right)
+ U_{2}\cos \left( 2x\right)\bigr] \psi (x)
= \tilde{E}\psi (x).
\end{equation}
Now we can utilize the translation operators 
\( \exp \left( \pm 2\pi \frac{\Phi _{0}}{\Phi }\partial_{x}\right)  \)
and \( \exp \left( \pm 4\pi \frac{\Phi _{0}}{\Phi }\partial_{x}\right),  \) 
yielding
\begin{equation}
2\left[
t_{1}\cos(-i2\pi \frac{\Phi _{0}}{\Phi }\partial_{x})
+t_{2}\cos(-i4\pi \frac{\Phi _{0}}{\Phi }\partial_{x})
+U_{1}\cos(x)
+U_{2}\cos(2x)
\right] \psi (x) = \tilde{E}\psi (x).
\end{equation}
\end{widetext}
Comparing this with the time-independent Schrödinger equation
gives the Hamiltonian,
\begin{equation}
\label{hamclass}
H=2\left[t_{1}\cos(p)+t_{2}\cos(2p)
+U_{1}\cos(x)+U_{2}\cos(2x) \right],
\end{equation}
with a momentum operator
\( p=-i\hbar_{\mathrm{eff}} \partial _{x} \) and
an effective Planck constant
\( \hbar _{\mathrm{eff}}=2\pi \frac{\Phi _{0}}{\Phi } \).
The corresponding classical Hamiltonian is one-dimensional and therefore
integrable.
The ubiquitous appearance of avoided band crossings,
as in chaotic systems, is thus not expected.

%%%%%%%%%%%%%%%%%% Figure 5 %%%%%%%%%%%%%%%%%%%%%%%%%%%
\begin{figure}
{\centering 
  \psfrag{x}[c]{$x$}\psfrag{y}[c]{$p$}
  \psfrag{-pi}[c]{$-\pi$}\psfrag{pi}[c]{$\pi$}
  \resizebox*{0.5\textwidth}{!}{
    \includegraphics{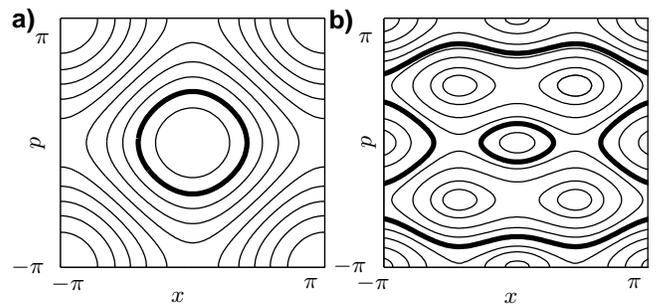}
  } \par 
}

\caption{\label{Abb:ClassicalHarper}
Iso-potential lines of the classical Hamiltionian Eq.~(\ref{hamclass}) with
(a) $t_2=U_2=0$
and (b) $t_2, U_2 \ne 0$. While in (a) there are no energy
degeneracies, in (b) several maxima and minima occur and 
energy degeneracies exist, e.g. the one depicted by thick lines.
}
\end{figure}
%%%%%%%%%%%%%%%%%%%%%%%%%%%%%%%%%%%%%%%%%%%%%%%%%%%%%%%

Still, we observe in Figs.~\ref{Abb:Harper2EMU1} and \ref{Abb:Harper2U1}
some avoided band crossings.
We explain this by studying the classical Hamiltonian Eq.~(\ref{hamclass}).
For the case of a purely electric modulation (Harper equation), $t_2=U_2=0$, it
has a single maximum and a single minimum within a unit cell.
If, however, $t_2$ and $U_2$ become non-zero, additional minima
or maxima appear and classically energetic degeneracies become possible, as
depicted in Fig.~\ref{Abb:ClassicalHarper}.
Quantum mechanically these are split by tunneling and lead to avoided crossings
under the variation of a parameter of the Hamiltonian.
The width of such avoided crossings, of course, depends on the
effective Planck constant $\hbar _{\mathrm{eff}}$.
Here we have studied values of $\hbar _{\mathrm{eff}}$, which are of the
order of 1, explaining the easily observable avoided crossings
of Figs.~\ref{Abb:Harper2EMU1} and \ref{Abb:Harper2U1}.
From the classical Hamiltonian Eq.~(\ref{hamclass})
we see that a magnetic modulation, which leads to $t_2, U_2 \ne 0$, is a
necessary condition for the appearance of classical energy degeneracies and
quantum avoided crossings.
We note, that if one goes beyond the lowest Fourier components of the 
spatial modulation, Eqs.~(\ref{PotLowFourier}) and (\ref{MagLowFourier}),
a purely electric modulation may also lead to energy degeneracies.

Now the usual reasoning applies~\cite{KarstenPaper}, 
namely that avoided band crossings lead
to metal-insulator transitions and the coexistence of localized
and extended states, explaining the observations in 
Figs.~\ref{Abb:Harper2EMU1} and \ref{Abb:Harper2U1}.
The dynamics of wave packets would show the corresponding changes in 
transport direction similar to Fig.~\ref{Abb:Spreading}.

\section{Conclusion}

We derive a matrix equation for the energy spectrum of electrons
in 2D periodic electric and magnetic fields.
We observe avoided band crossings, metal-insulator transitions,
and changes in the spreading direction of wave packets.
These phenomena, previously studied in the case without magnetic modulation,
are now found in two new situations.

Firstly, we find that they occur in a purely magnetic modulation.
The origin being the classically chaotic dynamics, which 
leads to avoided band crossings.
Secondly, we study the limit of weak electric and magnetic modulation, where
the Landau band coupling can be neglected.
For the energy spectrum we derive a 1D discrete Schr\"odinger equation 
with next-nearest neighbor coupling.
Its classical counterpart is shown to be an integrable 1D Hamiltonian
and naively one would not expect avoided band crossings.
We find, however, that avoided crossings exist and argue that they are due to
tunneling.
As a necessary condition for the appearance of avoided band crossings in this
limit we find either a nonzero magnetic modulation 
or a higher-order electric modulation.
These avoided band crossings lead to metal-insulator transitions and 
changes in the spreading direction of wave packets. 

\section{Acknoledgements}
We acknowledge fruitful discussions with T.~Geisel.

\end{document}